\documentclass [12pt]{iopart}
\usepackage{iopams}
\usepackage{graphicx}
\begin{document}
\title{Spatial and Wavenumber Resolution of Doppler Reflectometry}
\author{E Z Gusakov and A V Surkov}
\address{Ioffe Institute, Politekhnicheskaya 26,
194021 St. Petersburg, Russia}
\eads{\mailto{Evgeniy.Gusakov@mail.ioffe.ru},
\mailto{a.surkov@mail.ioffe.ru}}
\begin{abstract}
Doppler reflectometry spatial and wavenumber resolution is
analyzed within the framework of the linear Born approximation in
slab plasma model. Explicit expression for its signal
backscattering spectrum is obtained in terms of wavenumber and
frequency spectra of turbulence which is assumed to be radially
statistically inhomogeneous. Scattering efficiency for both back
and forward scattering (in radial direction) is introduced and
shown to be inverse proportional to the square of radial
wavenumber of the probing wave at the fluctuation location thus
making the spatial resolution of diagnostics sensitive to density
profile. It is shown that in case of forward scattering additional
localization can be provided  by the antenna diagram. It is
demonstrated that in case of backscattering the spatial resolution
can be better if the turbulence spectrum at high radial
wavenumbers is suppressed. The improvement of Doppler
reflectometry data localization by probing beam focusing onto the
cut-off is proposed and described. The possibility of Doppler
reflectometry data interpretation based on the obtained
expressions is shown.
\end{abstract}
\submitto{\PPCF}  \maketitle
\section{Introduction}
\label{sec:Intro} Plasma rotation velocity measurements are of
great importance for understanding transition to improved
confinement in tokamaks and physics of transport barriers.
Extensively used nowadays for such investigations is Doppler
reflectometry~\cite{Zou,Bulanin00,Hirsch01}. This technique
provides measuring fluctuations propagation poloidal velocity
which is often shown to be dominated by $\vec E \times \vec B$
velocity of plasma~\cite{Hirsch01}. In using this method a probing
microwave beam is launched into the plasma with finite tilt angle
with respect to density gradient. A back-scattered signal with
frequency differing from the probing one is registered by a nearby
standing or the same antenna. The information on plasma poloidal
rotation is obtained in this technique from the frequency shift of
the backscattering spectrum which is supposed to originate from
the Doppler effect due to the fluctuation rotation. Spatial
distribution of the scattering phenomena, which is usually assumed
to occur in the cut-off vicinity, is the key issue for the
diagnostic applications.

Some numerical simulations undertaken (see~\cite{Hirsch01} and
references there, \cite{Bulanin02}) and analytical
results~\cite{GusYakovlev} demonstrate the possibility of the
measurements localization by the cut-off but the problem of
Doppler reflectometry locality remains still open and lacks
comprehensive analytical treatment. This stems from the fact that
to be sure that main contribution to the registered signal is
made by the cut-off vicinity one should compare it with integral
contributions of distant from the cut-off regions which can be
substantial depending on density profile.

\begin{figure}
\begin{center}
\includegraphics[height=0.2\textheight]{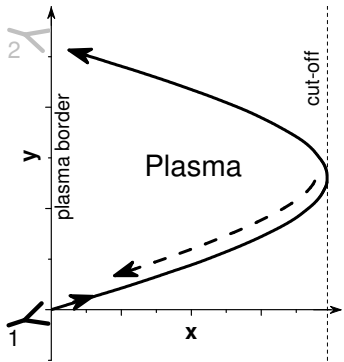}
\end{center}
\caption{\label{fig:Scheme} Diagnostics scheme. 1---emitting and
receiving antenna, 2---additional receiving antenna (to be
discussed in~\sref{sec:Disc}).}
\end{figure}

This paper attempts to clarify this problem. Here we present a
theoretical investigation of the Doppler reflectometry signal
dependence on the turbulence distribution with respect to the
cut-off. We consider slab two-dimensional model
(see~\fref{fig:Scheme}). This simplification allows us to perform
straightforward analytical treatment and obtain explicit reliable
expressions for the scattered signal which can be easily used for
estimation of diagnostics locality and experimental data
interpretation without complicated and time consuming numerical
calculation using full-wave codes, etc. The model considered can
be readily applied to large in comparison with probing beam width,
vertically elongated (ITER-like) plasma. The limitations of this
approximation are discussed below.

The paper is organized as follows. In \sref{sec:WKB} the
consideration is carried on in the geometrical optics
approximation for arbitrary plasma density profile. This
approximation fails to hold in the cut-off region. So in this
region the analysis is made assuming the density profile to be
linear and using exact expressions for the probing and scattered
wave electric field given by Airy functions (\sref{sec:Airy}).
Some numerical examples are given in \sref{sec:Num}. Brief
discussion of the model considered and results obtained is
provided in \sref{sec:Disc}. Finally, a conclusion follows in
\sref{sec:Concl}.
\section{Reflectometry signal in WKB--approximation}
\label{sec:WKB}
In this section the reflectometry signal is obtained in
geometrical optics (or WKB) approximation. The plasma is assumed
to be nonuniform in $x$ (radial) direction and uniform in $y$
(poloidal) direction. External magnetic field is supposed to be
along  $z$ axis. O-mode Doppler reflectometry is considered but
the final results can be easily adapted to X-mode reflectometry
applying corresponding expression for wave radial wavenumber
dependence on the radial coordinate.

A received signal is obtained using reciprocity
theorem~\cite{PiliyaPopov} and is assumed to be created by single
scattering (linear) mechanism. The frameworks of this
approximation and experimental means to check its applicability
are discussed in \sref{sec:Disc}. The scattering signal averaging
is made taking into account radial statistical inhomogeneity of
the turbulence. Then a scattering efficiency is introduced and
analyzed.
\subsection{Scattering signal}
We consider normalized antenna electric field in the following
form
\[
\vec E_a(\vec r)=\vec e_z \int\limits_{-\infty}^{+\infty}
\frac{\rmd k_y}{2\pi } W(x,k_y) f(k_y) \rme^{\rmi k_yy}
\]
where factor $f(k_y)$ takes into account the antenna pattern
describing antenna radiation in vacuum
\[
f(k_y)=\sqrt{\frac{c}{8\pi }} \int\limits_{-\infty}^{+\infty} dy
E_0(x=0,y) \rme^{-\rmi k_yy}
\]
The vacuum antenna field $E_0$ differs from $E_a$ by the absence
of the reflected wave contribution.

 Here and further we suppose the probing wave to oscillate at
frequency $\omega $ and omit the corresponding term in equations.
Radial electric field distribution is described by the function
$W(x,k_y)$ which is determined by
\begin{equation}
W''+k_x^2(x,k_y) W=0 \label{eq:W}
\end{equation}
where the square of ordinary wave wavevector radial projection is
given by $k_x^2(x,k_y)=k^2(x)-k_y^2=[\omega ^2-\omega
_{pe}^2(x)]/c^2 -k_y^2$. Thus ordinary wave electric field in
WKB-approximation has the following form~\cite{Tyntarev}:
\begin{eqnarray*}
\fl W(x,k_y)=4\sqrt{\frac{2\pi \omega }{c^2 k_x(x,k_y)} }
\exp\left[\rmi\int_0^{x_c(k_y)} k_x(x',k_y)\,\rmd x' -\frac{\rmi
\pi }{4} \right]  \\
\lo{\times} \cos\left[ \frac{\pi }{4} -\int_x^{x_c(k_y)}
k_x(x',k_y)\,\rmd x'\right]
\end{eqnarray*}
where the turning point $x_c(k_y)$ is determined by the equation
\[
k_x\left[x_c(k_y),k_y\right]=0
\]
and $x=0$ corresponds to the plasma border.

Using reciprocity theorem~\cite{PiliyaPopov, Tyntarev}, an
amplitude of the received scattering signal at frequency $\omega
_s=\omega +\Omega $, where $\Omega $ denotes the frequency of the
fluctuation caused the scattering, can be written as
\[
A_s(\omega _s)=\frac{\rmi e^2}{4 m_e\omega }\sqrt{P_i}
\int_{-\infty}^{+\infty} \delta n_{\Omega}(\vec r) E_a^2(\vec
r)\,\rmd\vec r
\]
Here $P_i$ is the probing wave power.  Following the model
considered the fluctuations are assumed to be long enough along
the magnetic field direction $z$, so their dependence on $z$ can
be neglected. Introducing the density perturbation Fourier
harmonic
\[
\delta n(\varkappa,q,\Omega )=\int_{-\infty}^{+\infty} \delta
n(x,y,\Omega ) \rme^{-\rmi \varkappa x-\rmi qy} \rmd x \rmd y
\]
we obtain
\begin{equation}
\fl A_s(\omega _s)=\frac{\rmi \pi e^2}{2 m_e\omega }\sqrt{P_i}
\int_{-\infty}^{+\infty} \frac{\rmd k_y\,\rmd\varkappa\,\rmd
q}{(2\pi )^3}\, \delta n(\varkappa,q,\Omega )
 f(k_y) f(-k_y-q) C(\varkappa ,q,k_y)
 \label{eq:A:C}
\end{equation}
 The efficiency of scattering
off the fluctuation with radial wavenumber $\varkappa$ and
poloidal wavenumber $q$ has the following form~\cite{Tyntarev}
\begin{eqnarray*}
\fl C(\varkappa ,q,k_y)=\int_{-\infty}^{+\infty}
W(x,k_y)W(x,-k_y-q)\rme^{\rmi \varkappa x}\rmd x\\
\lo{=}\int_{-\infty}^{+\infty} \frac{\rmd x}{
\sqrt{k_x(x,k_y)k_x(x,-k_y-q)}}
\sum_{m,n=\pm 1} \rme^{\rmi \Psi _{mn}-\rmi(m+n)\pi /4} \\
 \Psi _{mn}=\varkappa
x+m\phi(x,k_y)+n\phi(x,-k_y-q)+\phi(0,k_y)+\phi(0,-k_y-q)\\
\phi(x,k_y)=\int_x^{x_c(k_y)} k_x(x',k_y)\,\rmd x'
\end{eqnarray*}
The scattering efficiency $C(\varkappa ,q,k_y)$ is an integral of
the oscillating function. Main contribution to this integral is
made by stationary phase points. Following~\cite{Tyntarev} to
estimate this integral let us calculate this contribution.
Equalizing the derivative of the phase to zero one obtains the
Bragg condition
\begin{equation}
\varkappa -mk_x(x_*,k_y)-nk_x(x_*,-k_y-q)=0 \label{eq:Bragg}
\end{equation}
In nonuniform plasma this determines the scattering point $x_*$
\begin{equation}
k^2(x_*)=\left[\frac{\varkappa }{2}+\frac{q(2k_y+q)}{2\varkappa
}\right]^2 +k_y^2 \label{eq:xstar}
\end{equation}
The figures $m, n$ are related to different cases of scattering.
\begin{equation}
m={\rm sgn} \frac{\varkappa ^2+q(2k_y+q)}{\varkappa }, \quad
n={\rm sgn} \frac{\varkappa ^2-q(2k_y+q)}{\varkappa }
\label{eq:mn}
\end{equation}
As it will be shown below case $m=n$ corresponds to the
backscattering (BS) and $m=-n$ accords to the forward scattering
(FS). The sign of $m$ is related to the scattering before $(m=-1)$
and after $(m=1)$ the cut-off in respect of the probing wave
propagation. Equations~(\ref{eq:mn}) express the fact that one
fluctuation can scatter the wave only once. The situation, for
example, when the wave with fixed $k_y$ is forward scattered off
the fluctuation far from the cut-off and the same wave can be
scattered backward off the same fluctuation near the cut-off is
impossible. This circumstance will allow us below to separate the
contributions of forward and backward scattering processes.

The final expression for the scattering efficiency in WKB
approximation takes the form
\[
C(\varkappa ,q,k_y)=2\sqrt{\frac{\pi\ell_*^3}{|\varkappa|}}
\exp\left[\rmi\Psi_{mn}(x_*)+\frac{\rmi\pi}{4} {\rm sgn}(
mn\varkappa)\right]
\]
where $\ell_*=\left[\partial k^2(x)/\partial
x|_{x=x_*}\right]^{-1/3}$ is local Airy scale. Introducing local
density variation scale $L_*=[\rmd\ln n_e(x)/\rmd
x|_{x=x_*}]^{-1}$  we obtain for ordinary probing wave $\ell
_*=\left[c^2 L_*/\omega _{pe}^2(x_*)\right]^{1/3}$.

 On substituting the obtained expression for the scattering
efficiency into~\eref{eq:A:C} we obtain the received signal
\begin{eqnarray}
\fl A_s=4\pi ^{3/2}\frac{e^2}{m_e c^2 }\sqrt{P_i}
\int_{-\infty}^{+\infty} \frac{\rmd k_y\,\rmd\varkappa\,\rmd
q}{(2\pi )^3}\, \delta n(\varkappa ,q,\Omega )
 f(k_y) f(-k_y-q) \frac{\ell _*^{3/2}}{\sqrt{\varkappa
-\rmi o}}\,\rme^{\rmi \Psi(x_*,k_y)}
 \label{eq:A:ky}
\end{eqnarray}
where $o>0$ determines how the integration path goes around the
singularity.  Here $\Psi (x_*,k_y)=\Psi _{mn}$, where $m, n$ is
determined by \eref{eq:mn}.

\subsection{Scattering signal analysis: integration over $k_y$}
\label{sec:ky} We calculate the integral over $k_y$
in~\eref{eq:A:ky} using saddle point method. It is especially
efficient if the cut-off is far enough from the antenna and does
not coincide with the focal point of the antenna beam, so that the
ray tracing consideration is applicable. In this case the saddle
point $k_y^*$ is determined by the stationary point of the phase
$\Phi=\Psi(x_*,k_y)+\arg f(k_y)+\arg f(-k_y-q)$ and the
$|f(k_y)f(-k_y-q)|$ dependence on $k_y$ is negligible.
Corresponding criterion to distinguish this and opposite case will
be formulated below.

In this ``ray tracing'' case the stationary phase condition
\begin{equation}
\left.\frac{\rmd\Phi \left[ x_*(k_y),k_y\right]}{\rmd k_y}
\right|_{k_y=k_y^*}=0 \label{eq:stph:ky}
\end{equation}
can be easily interpreted. At first we consider the meaning of
$\rmd\Psi/\rmd k_y$ term. Recollecting that
\begin{equation}
k_x^2(x,k_y)=k^2(x)-k_y^2 \label{eq:k2}
\end{equation}
and
\[
\frac{\partial\phi(x,k_y)}{\partial
k_y}=-k_y\int_x^{x_c(k_y)}\frac{\rmd x'}{k_x(x',k_y)}
\]
one obtains
\begin{eqnarray}
\fl\frac{\partial\Psi}{\partial
k_y}=-mk_y\int_{x_*}^{x_c(k_y)}\frac{\rmd
x}{k_x(x,k_y)}-n(k_y+q)\int_{x_*}^{x_c(-k_y-q)}\frac{\rmd
x}{k_x(x,-k_y-q)}\nonumber\\
\lo{-}k_y\int_{0}^{x_c(k_y)}\frac{\rmd
x}{k_x(x,k_y)}-(k_y+q)\int_{0}^{x_c(-k_y-q)}\frac{\rmd
x}{k_x(x,-k_y-q)} \label{eq:dpsi:dky}
\end{eqnarray}
Taking into account that \eref{eq:k2} yields the following
relation between the projections of group velocities $\vec
v_g=\partial\omega/\partial\vec k$ of incident and scattered
waves:
\[
\frac{v_{gy}^{(i)}}{v_{gx}^{(i)}}=\frac{k_y}{k_x(x,k_y)}, \qquad
\frac{v_{gy}^{(s)}}{v_{gx}^{(s)}}=\frac{-k_y-q}{k_x(x,-k_y-q)}
\]
we obtain
\begin{equation}
\fl\frac{\partial\Psi}{\partial
k_y}=-m\int_{x_*}^{x_c(k_y)}\frac{v_{gy}^{(i)}}{v_{gx}^{(i)}}\,\rmd
x-\int_{0}^{x_c(k_y)}\frac{v_{gy}^{(i)}}{v_{gx}^{(i)}}\,\rmd x
+n\int_{x_*}^{x_c(-k_y-q)}\frac{v_{gy}^{(s)}}{v_{gx}^{(s)}}\,\rmd
x+\int_{0}^{x_c(-k_y-q)}\frac{v_{gy}^{(s)}}{v_{gx}^{(s)}}\,\rmd x
\label{eq:psi:ky}
\end{equation}
Now one can see that $\Delta y=-\partial\Psi/\partial k_y$
corresponds to shift of the ray trajectory along $y$ direction
when it returns to the antenna. In less general case it was
mentioned in~\cite{GusYakovlevPPCF}.
 For instance, if $m=-n=1$ then \eref{eq:psi:ky} takes the
form
\begin{equation*}
\frac{\partial\Psi}{\partial
k_y}=-\int_{0}^{x_c(k_y)}\frac{v_{gy}^{(i)}}{v_{gx}^{(i)}}\,\rmd
x-\int_{x_c(k_y)}^{x_*}\frac{v_{gy}^{(i)}}{-v_{gx}^{(i)}}\,\rmd x
-\int_{x_*}^{0}\frac{v_{gy}^{(s)}}{v_{gx}^{(s)}}\,\rmd x
\end{equation*}
and $\left(-\partial\Psi/\partial k_y\right)$ is the shift of the
ray trajectory in case of forward scattering after the reflection
off the turning point. Here the meaning of the figures
$m,n$~(\ref{eq:mn}) announced above becomes clear.

Accounting for the influence of the wavefront curvature at the
antenna given by $\arg f(k_y)\neq0$ we consider gaussian antenna
pattern
\begin{equation}
f(k_y)=\sqrt{2\sqrt{\pi}\rho}\rme^{-(\rho ^2-\rmi
c\mathcal{R}/\omega)(k_y-\mathcal{K})^2/2} \label{eq:gauss}
\end{equation}
In case of
\begin{equation}
c\mathcal{R}/\omega\gg\rho^2\label{eq:condR}
\end{equation}
{parameter $\mathcal{R}$ has a meaning of a wavefront curvature
radius at the antenna.} In~\eref{eq:gauss} $\mathcal{K}$
corresponds to the antenna tilt
($\mathcal{K}=\omega/c\sin\vartheta$, where $\vartheta$ denotes
tilt angle in respect of the density gradient). This allows us to
obtain condition~(\ref{eq:stph:ky}) in the form
\begin{equation}
\frac{\rmd\Phi}{\rmd k_y}=-\Delta
y+\frac{c\mathcal{R}}{\omega}(2k_y+q)=0 \label{eq:stph:ky:m}
\end{equation}
Taking into account that the ray with poloidal wavevector
component $k_y$ is radiated from the position
$y(k_y)=-k_y\mathcal{R}c/\omega$ of the curved wavefront at the
antenna which is assumed to be situated in the axes origin, it is
easy to show that~\eref{eq:stph:ky:m} determines a ray trajectory
which returns to the proper point of the antenna wavefront after
the scattering off the fluctuation with wavevector~$(\varkappa
,q)$.

To get explicit expression for the stationary point position
$k_y^*$ we use paraxial approximation. Supposition of small
divergence of the antenna beam $|k_y-\mathcal{K}|\ll \omega/c$
which holds usually in the experiments on Doppler
reflectometry~\cite{Hirsch01} allows us to write
\[
k_x(x,k_y)\approx
k_x(x,\mathcal{K})-\frac{k_y^2-\mathcal{K}^2}{2k_x(x,\mathcal{K})}
\]
and neglect the dependence on $k_y$ in denominators
in~\eref{eq:dpsi:dky}. This allows us to obtain
\begin{eqnarray*}
\frac{\partial\Psi}{\partial
k_y}=-\frac{ck_y}{\omega}\left[(m+n)\Lambda(x_*)
+2\Lambda_0\right]-\frac{cq}{\omega} \left[n\Lambda(x_*)
+\Lambda_0\right]\\
\Lambda(x)=\frac{\omega }{c}\int_x^{x_c(\mathcal{K})} \frac{\rmd
x'}{k_x(x',\mathcal{K})},\quad \Lambda_0\equiv \Lambda(0)
\end{eqnarray*}

Using~\eref{eq:stph:ky} we get the stationary phase point position
\begin{equation}
 k_y^*=\cases {
-q/2,&m=n \\  -q/2\left[1+n \Lambda
(x_*)\left/(\Lambda_0-\mathcal{R})\right.\right],&m=-n }
\label{eq:kystar}
\end{equation}
for the BS and FS respectively.

In the opposite case when ray trajectories consideration is not
valid  $k_y^*$ is determined by the antenna pattern amplitude
\[
\left.\frac{\rmd|f(k_y)f(-k_y-q)|}{\rmd k_y}\right|_{k_y=k_y^*}=0
\]
yielding $k_y^*=-q/2$ for an arbitrary antenna pattern.

The assumption of gaussian antenna pattern~(\ref{eq:gauss}) allows
us to evaluate the saddle point position in general case
\begin{equation*}
k_y^*=-\frac{q}{2}\frac{\Lambda_0-\mathcal{R}+n\Lambda(x_*)-\rmi\mathcal{P}}{
\Lambda_0-\mathcal{R}+(m+n)/2\cdot\Lambda(x_*)-\rmi\mathcal{P}}
\end{equation*}
where $\mathcal{P}=\omega\rho^2/c$ and formulate the criterion in
question. The ray tracing consideration is valid if $\rho^2\ll
c|\Lambda_0-\mathcal{R}|/\omega$ and thus the focal point is not
too close to the cut-off. If the focal point is situated in the
cut-off $\rho^2\gg c|\Lambda_0-\mathcal{R}|/\omega$ we come to the
expression $k_y^*=-q/2$.

Using stationary $k_y^*$ calculated above and performing the
integration we obtain the scattering signal in the following form
\begin{eqnarray*}
\fl A_s=2\pi\frac{e^2}{m_e c^2 } \sqrt{P_i}
\int_{-\infty}^{+\infty} \frac{\rmd\varkappa\,\rmd q}{(2\pi )^2}
\, \delta n(\varkappa ,q,\Omega ) \\
\lo{\times} f\left[k_y^*(\varkappa ,q)\right]
f\left[-k_y^*(\varkappa
 ,q)-q\right]
 \ell _*^{3/2} \frac{\Delta (\varkappa, q)}{\sqrt{\varkappa -\rmi o}}
 \rme^{\rmi \Psi(x_*,k_y^*)}
\end{eqnarray*}
where
\[
\Delta (\varkappa, q)=\left\{\rho
^2-\rmi\left[\frac{c\mathcal{R}}{\omega}+\frac12\frac{d^2\Psi
\left[ x_*(k_y),k_y\right]}{\rmd k_y^2}\right]\right\}^{-1/2}
\]
\subsection{Scattering signal averaging}
We consider the turbulence to be slightly inhomogeneous along $x$
direction so that the density fluctuation correlation function
takes the form
\begin{equation}
\fl\langle\delta n (x) \delta n (x')\rangle=\delta
n^2\left(\frac{x+x'}{2}\right)
 \int_{-\infty}^{+\infty} \frac{\rmd\varkappa }{2\pi }
\left|\tilde
n\left(\varkappa,q,\Omega,\frac{x+x'}{2}\right)\right|^2\rme^{\rmi
\varkappa (x-x')} \label{eq:inhom}
\end{equation}
This representation is applicable when the turbulence correlation
length along $x$ axis $\ell_{cx}$ is much smaller than the
turbulence inhomogeneity scale. It allows us to take into account
the dependence of turbulence on the radial coordinate and still
describe it using wavenumber spectrum $\left|\tilde
n\left[\varkappa,q,\Omega,(x+x')/2\right]\right|^2$. Supposing
the turbulence to be stationary and homogeneous in $y$-direction
and using \eref{eq:inhom} we can represent correlation function
of spectral density in the form
\begin{eqnarray*}
\fl \left\langle \delta n(\varkappa,q,\Omega)\delta
n^*(\varkappa',q',\Omega')\right\rangle=\int_{-\infty}^{+\infty}\rmd
x\,\delta n^2(x)\left|\tilde
n\left(\frac{\varkappa+\varkappa'}{2},q,\Omega,x\right)\right|^2\\
\lo{\times}\rme^{\rmi
x(\varkappa-\varkappa')}(2\pi)^2\delta(q-q')\delta(\Omega-\Omega')
\end{eqnarray*}
where the integration is performed over all plasma volume. It
allows spectral power density of the received signal to be
represented in the following form
\begin{equation}
 p(\omega _s)=\langle
A_s\bar{A_s}\rangle=P_i\int_{-\infty}^{+\infty} \rmd x \,\delta
n^2(x)S(x) \label{eq:ps}
\end{equation}
where $\bar{A_s}$ is complex conjugate to $A_s$.

 The scattering efficiency $S(x)$ introduced here takes the
form
\begin{eqnarray*}
\fl S(x)=4\pi^2\left(\frac{e^2}{m_e c^2 }\right)^2
\int_{-\infty}^{+\infty}
\frac{\rmd\varkappa\,\rmd\varkappa'\,\rmd q}{(2\pi )^3}
\, \left|n\left(\frac{\varkappa+\varkappa'}{2},q,\Omega,x\right)\right|^2\\
\lo{\times} f\left[k_y^*(\varkappa ,q)\right]
f\left[-k_y^*(\varkappa,q)-q\right]\overline{f\left[k_y^*(\varkappa'
,q)\right]}\,\overline{f\left[-k_y^*(\varkappa',q)-q\right]}\\
\times(\ell_*\ell_*')^{3/2} \frac{\Delta(\varkappa,
q)\Delta(\varkappa',q)}{\sqrt{\varkappa\varkappa'}}\,
 \rme^{\rmi x(\varkappa-\varkappa')+ \rmi\Psi[\varkappa,q]-\rmi\Psi[\varkappa',q]}
\end{eqnarray*}
and determines actually the Doppler reflectometry locality and
wavenumber resolution. Performing the integration over
$\varkappa-\varkappa'$ using stationary phase method we obtain
\begin{eqnarray}
\fl S(x)\approx (2\pi)^{3/2} \left(\frac{e^2}{m_e c^2 }\right)^2
\int_{-\infty}^{+\infty} \frac{\rmd\varkappa\,\rmd q}{(2\pi )^2}
\left|\tilde n(\varkappa ,q,\Omega,x)\right|^2
 \left|f\left[k_y^*(\varkappa ,q)\right]\right|^2
\left|f\left[-k_y^*(\varkappa ,q)-q\right]\right|^2  \nonumber\\
\lo{\times}
 \ell _*^3 \frac{\left|\Delta (\varkappa,
q)\right|^2}{\left|\varkappa
\right|\sqrt{\left|x_{*\varkappa}'\right|}} \exp\left\{\frac{\rmi
\left[x-x_*(\varkappa ,q)\right]^2}{2x_{*\varkappa }'(\varkappa
,q)} -\frac{\rmi \pi }{4} {\rm sgn} x_{*\varkappa }'(\varkappa
,q)\right\}\label{eq:Sx:osc}
\end{eqnarray}
where $x_{*\varkappa }'(\varkappa ,q)\equiv\partial x_*(\varkappa
,q)/\partial\varkappa$ and the scattering point position
$x_*(\varkappa ,q)$ is given by~\eref{eq:Bragg}.

One can see that due to the oscillating term $\exp\left\{\rmi
\left[x-x_*(\varkappa ,q)\right]^2/\left(2x_{*\varkappa
}'(\varkappa ,q)\right)\right\}$ the main input to the
integral~(\ref{eq:Sx:osc}) is provided by the stationary point
$x=x_*$. It corresponds to the contribution of the fluctuations
producing scattering just in the point $x$.

Here we suppose the fluctuation spectral density $\left|\tilde
n(\varkappa ,q,\Omega,x)\right|^2$ to vary with $\varkappa$ slow
enough in comparison with the oscillating term. In general case
four terms arise from condition $x_*(\varkappa ,q)=x$ which is
equivalent to
\begin{equation}
k\left[x_*(\varkappa^*, q)\right]=k(x) \label{eq:keqk}
\end{equation}
where $k(x_*)$ is determined by~\eref{eq:xstar} and $k_y=k_y^*$ is
substituted according to~\sref{sec:ky}. Solving~\eref{eq:keqk} and
taking interest in ``stationary'' $\varkappa^*$ we obtain
\begin{equation}
\varkappa_{m,n}^*=mk_x(x,k_y^*)+n\sqrt{k_x^2(x,k_y^*)-q(2k_y^*+q)}
\label{eq:solution}
\end{equation}
Figures $m$ and $n$ have the same meaning as above but here and
further they are independent parameters and specify corresponding
solutions of~\eref{eq:keqk}.

To distinguish contributions of different scattering types we
represent the scattering efficiency in the following form
\begin{equation}
 S(x)=\pi
\left(\frac{e^2}{m_e c^2 }\right)^2 \int_{-\infty}^{+\infty}
\frac{\rmd q}{2\pi }\left[ S_{BS}(x,q)+S_{FS}(x,q)\right]
\label{eq:Sx}
\end{equation}
and discuss the properties of back and forward scattering
efficiency separately.
\subsection{BS efficiency}
To obtain the two terms corresponding to the backscattering which
occur before or after reflection of the probing wave we substitute
$m=n$ and $k_y^*=-q/2$ to~\eref{eq:solution} according
to~\eref{eq:kystar}. This yields
\[
\varkappa_{m,m}^*=2mk_x\left(x,-\frac{q}{2}\right),\quad m=\pm1
\]
and
\begin{eqnarray*}
\left|x_{*\varkappa}'\right|=\ell_*^3k_x\left(x,-\frac{q}{2}\right)\\
\left|\Delta (\varkappa_{m,m}^*,
q)\right|^2=\left\{\rho^4+\frac{c^2}{\omega^2}\left[
\Lambda_0-\mathcal{R}+m\Lambda(x)\right]^2\right\}^{-1/2}
\end{eqnarray*}
It allows  us to obtain BS contribution in question
\begin{equation}
S_{BS}(x,q)=
\frac{\left|f\left(-q/2\right)\right|^4}{k_x^2\left(x,\mathcal{K}\right)}
\sum_{m=\pm 1}\frac{\left|\tilde
n\left[2mk_x\left(x,\mathcal{K}\right),q,\Omega,x\right]\right|^2}{\sqrt{\rho
^4+c^2\left[\Lambda_0-\mathcal{R}+m\Lambda(x)\right]^2/\omega ^2}}
\label{eq:SBS}
\end{equation}
It can be seen that the contributions of the BS before (m=n=-1)
and after (m=n=1) the cut-off have different amplitudes due to
the diffraction effect dependence on length of the ray trajectory
from emitting to receiving antenna.

We consider main features of the BS efficiency~(\ref{eq:SBS})
obtained. The first factor determining the localization of BS
signal which appears to be typical for the fluctuation
reflectometry~\cite{GusPopovEPS} is $k^{-2}(x)$. Just
recently~\cite{GusPopovEPS} the same factor was revealed in
analysis of correlation matrix of phase perturbations of
fluctuation reflectometry signal in nonlinear regime. It is
maximal in the vicinity of the cut-off where in the WKB
approximation it has singularity. This singularity saturation will
be treated in details in \sref{sec:Airy}. This maximum explained
by the growth of the probing and scattered wave electric filed in
the cut-off provides the technique with spatial localization,
however, the decay of $k^{-2}(x)$ when leaving cut-off is not fast
enough to guarantee suppression of the signal coming from wide
edge region. Moreover, for quite a few density profiles (e.g.
linear and bent down ones) the integral of $k^{-2}(x)$ over $x$
does not converge due to far from the cut-off regions. So plasma
periphery contribution can be essential in these cases. The
illustration of this effect is given in \sref{sec:Num}.

Additional possibility to increase selectively the BS signal
coming from the cut-off  can be provided by focusing the beam of
probing and receiving  antennae to the cut-off region. This
corresponds to $\mathcal{R}=\Lambda_0$ in~\eref{eq:SBS}.  In this
case the focusing causes additional growth at the cut-off of
probing and scattered wave amplitudes leading to the scattered
signal enhancement similar to that predicted for the cut-off or
upper hybrid resonance~\cite{Novik}.

The two mentioned effects  increasing the probing wave electric
field substantially  in the cut-off should enhance the locality of
the Doppler reflectometry diagnostics making it less sensitive to
the backscattering in the edge plasma.

The  backscattering locality can be also better for some
fluctuation radial wavenumber spectra. One can see
from~\eref{eq:SBS} that if the short-scale fluctuations are
suppressed enough in the spectrum the backscattering contribution
will be essential in the cut-off vicinity only.
\subsection{FS efficiency}
To calculate the FS terms we use ray-tracing consideration.
Substituting $m=-n$ to~\eref{eq:solution} we obtain
\[
\varkappa_{m,-m}^*=\frac{q^2\Lambda(x)}{2k_x(x,k_y^*)(\Lambda_0-\mathcal{R})}
\]
It is noteworthy that both $\varkappa_{m,-m}^*$ have the same sign
and differ only slightly due to different values of $k_y^*$ only.

Finally we obtain
\begin{eqnarray*}
\left|x_{*\varkappa}'\right|=
\frac{2c(\Lambda_0-\mathcal{R})}{\omega q^2}
k_x(x,k_y^*)k_x(x,-k_y^*-q)\left|\frac{\omega\Lambda(x)}{2cL_*k_x(x,k_y^*)}-1\right|^{-1}
\\
\left|\Delta (\varkappa_{m,-m}^*,
q)\right|^2=\left(\rho^4+\frac{c^2(\Lambda_0-\mathcal{R})^2}{\omega^2}\right)^{-1/2}
\left|1-\frac{2cL_*k_x(x,k_y^*)}{\omega\Lambda(x)}\right|^{-1}
\end{eqnarray*}

Thus the sum of two similar forward scattering terms corresponding
to the scattering before and after the turning point takes a form
\begin{eqnarray}
\fl S_{FS}(x,q)= \frac{2}{\sqrt{\rho
^4+c^2(\Lambda_0-\mathcal{R})^2/\omega^2}}
\left|f\left\{-\frac{q}{2}\left[1+\frac{\Lambda(x)}{\Lambda_0-\mathcal{R}}\right]\right\}
\right|^2\nonumber\\
 \lo{\times}
\left|f\left\{-\frac{q}{2}\left[1-\frac{\Lambda(x)}{\Lambda_0-\mathcal{R}}\right]\right\}\right|^2\frac{1}{k_x^2\left(x,\mathcal{K}\right)}\left|\tilde
n\left[\frac{q^2 \Lambda(x)}{2k(x)(\Lambda_0-\mathcal{R})}
,q,\Omega,x\right]\right|^2 \label{eq:SFS}
\end{eqnarray}
The main reason for the mentioned similarity is equal trajectory
length in this case. It is worth noting that if
$\mathcal{R}<\Lambda_0$ the forward scattering efficiency $S_{FS}$
is contributed to by the fluctuations with positive radial
wavenumbers only.

Substituting gaussian antenna power diagram~(\ref{eq:gauss})
into~\eref{eq:SFS} we obtain expression for the factor describing
antenna pattern influence in the following form
\begin{eqnarray}
\fl
\left|f\left[-\frac{q}{2}\left(1+\frac{\Lambda(x)}{\Lambda_0-\mathcal{R}}\right)\right]\right|^2
\left|f\left[-\frac{q}{2}\left(1-\frac{\Lambda(x)}{\Lambda_0-\mathcal{R}}\right)\right]\right|^2
\nonumber\\
\lo{=}4\pi\rho^2\exp\left\{-\frac{\rho^2}{2}\left[\left(q+2\mathcal{K}\right)^2+
\left[\frac{q\Lambda(x)}{\Lambda_0-\mathcal{R}}\right]^2\right]\right\}
\label{eq:ffgauss}
\end{eqnarray}

Additional contribution to the forward scattering efficiency can
be provided by singularities of the term $\left|\Delta (\varkappa,
q)\right|^2/\left|\varkappa \right|$ in~\eref{eq:Sx:osc}. As it is
shown in~\ref{sec:App} it is small at substantial distance from
the cut-off where the condition
\begin{equation}
\frac{ck(x)}{\omega}>\rho\sqrt\frac{\omega}{c(\Lambda_0-\mathcal{R})}
\left(\frac{12}{\alpha}\right)^{1/2} \label{eq:crit:branch}
\end{equation}
holds. Here
$\alpha=\left.\left|L_*^2/n_e(x_c)\cdot\rmd^2n_e(x)/\rmd
x^2\right|\right|_{x=x_c}$ characterizes the nonlinearity  of the
density profile. If the cut-off is not close to the focal point,
so that $\rho\sqrt{\omega/[c(\Lambda_0-\mathcal{R})]}\ll 1$ the
expression~(\ref{eq:SFS}) for the FS contribution appears to be
valid where nonlinear corrections to $k^2(x)$ are still small.
Closer to  the  cut-off where inequality~(\ref{eq:crit:branch}) is
not satisfied the contribution of the branching point to the
integral in~\eref{eq:Sx:osc} is important. It is calculated below
taking into account that the density profile in this region can be
treated as linear. The corresponding contribution doubles the
result for the FS signal.

Similar to~\eref{eq:SBS} FS efficiency~(\ref{eq:SFS}) is also
proportional to $k^{-2}(x)$. However unlike~\eref{eq:SBS} the
cut-off contribution is not enhanced there by probing wave
focusing. In this case the focusing merely compensates the
refraction of probing beam leading to FS signal amplitude growth
in all plasma volume. It can be seen from~\eref{eq:SFS} that FS
signal is contributed to by long-scale fluctuations, which disable
the third possible localizing factor discussed for BS ---
turbulence spectrum.

Nevertheless, according to~\eref{eq:SFS},~\eref{eq:ffgauss} the
extra localization of the forward scattering can be due to the
fact that FS signal coming from plasma volume is received in
Doppler reflectometry is received by the antenna pattern
periphery. Supposing gaussian antenna beam to be wide enough, the
integration over $q$ yields the following estimation of the
forward scattering efficiency
\begin{eqnarray*}
\fl S_{FS}(x)=\pi \left(\frac{e^2}{m_e c^2 }\right)^2
\int_{-\infty}^{+\infty} \frac{\rmd q}{2\pi } S_{FS}(x,q)\nonumber\\
\lo{\approx}\left(\frac{e^2}{m_e c^2 }\right)^2
\frac{2\left(2\pi\right)^{3/2}\omega\rho}{c(\Lambda_0-\mathcal{R})
k^2_x(x)} \exp\left\{-2\left[\frac{\rho
\mathcal{K}\Lambda(x)}{\Lambda_0-\mathcal{R}}\right]^2\right\}\nonumber\\
\times\left|\tilde n\left[\frac{2\mathcal{K}^2 \Lambda
(x)}{k(x)(\Lambda_0-\mathcal{R})}
,-2\mathcal{K},\Omega,x\right]\right|^2
\end{eqnarray*}
Thus the factor describing the antenna pattern effect decreases
rapidly in moving off the cut-off under condition the beam is wide
enough or sufficiently tilted. If the antenna beam focusing to the
cut-off is provided this localizing factor can be even more
essential.
\section{Scattering signal in the cut-off vicinity}
\label{sec:Airy}
The explicit expressions for the scattering
efficiency~(\ref{eq:SBS}),~(\ref{eq:SFS}) were obtained in the
previous section using WKB and ray tracing consideration. These
expressions possesses singularities when validity conditions for
WKB approach
\begin{equation}
k^{-2}(x)\frac{\rmd k(x)}{\rmd x}\ll 1 \label{eq:WKB:crit}
\end{equation}
and ray tracing consideration
\begin{equation*}
\rho^2\ll \left|\frac{c(\Lambda_0-\mathcal{R})}{\omega}\right|
\end{equation*}
are violated in the cut-off vicinity.

To analyze the scattering efficiency in this region a more
rigorous approach is needed. To do that we recollect that in the
cut-off vicinity the density profile can be represented as linear
$k^2(x)=(x_c-x)/\ell ^3$ where $\ell =\left(c^2 L/\omega
^2\right)^{1/3}$ is the Airy scale and $L=[\rmd\ln n_e(x)/\rmd
x|_{x=x_c}]^{-1}$ is local density variation scale in the cut-off
position. In this case the criterion~(\ref{eq:WKB:crit}) takes the
form $k(x)\ell\gg 1$.

 Then in the cut-off vicinity the radial distribution of the
ordinary wave electric field has the following
form~\cite{Tyntarev}:
\begin{eqnarray*}
\fl W(x,k_y)=\sqrt{\frac{8 \omega \ell }{c^2}}
 \exp\left[\rmi\int_0^{x_c(k_y)} k_x(x',k_y)\,\rmd x'
-\frac{\rmi \pi }{4}\right]\\ \lo{\times}
\int_{-\infty}^{+\infty}\exp\left[\frac{\rmi
p^3}{3}+(\xi+k_y^2\ell^2)p\right]\,\rmd p
\end{eqnarray*}
where $ \xi=(x-x_c)/\ell$.

According to the reciprocity theorem~\cite{PiliyaPopov} the
scattering signal has the following structure
\begin{eqnarray}
\fl A_s=\frac{\rmi e^2}{4 m_e\omega }\sqrt{P_i} \int \rmd x
\int_{-\infty}^{+\infty} \frac{\rmd k_y\,\rmd\varkappa\,\rmd
q}{(2\pi )^3} \, \delta n(\varkappa,q,\Omega )
 \rme^{\rmi \varkappa (x-x_c)}\nonumber\\
 \lo{\times} f(k_y) f(-k_y-q) W(x,k_y) W(x,-k_y-q)
 \label{eq:As:Airy}
\end{eqnarray}
As we have already done assuming moderate spatial inhomogeneity of
the turbulence~(\ref{eq:inhom}) and calculating the integral
over~$k_y$ in~\eref{eq:As:Airy} by stationary phase method we
transform the registered signal spectral power density to the
form~(\ref{eq:ps}). For the scattering efficiency we get
\begin{eqnarray}
\fl S(x)\approx
\frac{4\pi\ell^3}{\sqrt{\rho^4+c^2(\Lambda_0-\mathcal{R})^2/\omega
^2}}\left(\frac{e^2}{m_e c^2 }\right)^2\nonumber\\
\lo{\times}\int_{-\infty}^{+\infty} \frac{\rmd\varkappa\,\rmd
q}{(2\pi )^2} \left|\tilde n(\varkappa ,q,\Omega,x)\right|^2
 \left|f\left(-\frac{q}{2}\right)\right|^4 R(x,\varkappa ,q) \label{eq:Sx:COL}\\
\fl R(x,\varkappa ,q)=\int_{-\infty}^{+\infty} \frac{\rmd\tau
}{\sqrt{\left(\beta -\theta\right)^2-\left(\tau
+\rmi\epsilon\right)^2}}\exp{\left\{ \frac{\rmi \tau
^3}{6}+\frac{\rmi \tau }{2} \left(\beta ^2 -K^2\right) \right\}}
\end{eqnarray}
where the following notation is used
\[
\fl\beta =\varkappa \ell,\quad\theta=\frac{L\ell
cq^2}{\omega}\frac{\Lambda_0-\mathcal{R}}{(\Lambda_0-\mathcal{R})^2+
\mathcal{P}^2},\quad\epsilon
=\frac{\mathcal{P}\theta}{\Lambda_0-\mathcal{R}},\quad K=2\ell
k_x\left(x,-\frac{q}{2}\right)
\]
The expression for $R(x,\varkappa ,q)$ can be simplified in two
cases. In the first one $K\gg \theta $ and the position of the
turbulence $x$ is far from the cut-off. Omitting oscillating terms
hardly having an effect in integrating over $\beta $ we can
distinguish two characteristic ranges of fluctuation radial
wavenumbers. The fist one corresponds to $|\beta |\ll K$.
Considering such fluctuations, which are responsible for the
forward scattering we can approximate
\begin{equation}
R\approx R^{(1)}(x,\varkappa ,q)=2\pi {\rm J_0}\left[ \frac12
K^2\left|\beta -\theta \right| \right] \rme^{-\epsilon K^2}
\label{eq:R1}
\end{equation}
The second group of wavenumbers corresponds to $|\beta |\sim K$.
These fluctuations provide the backscattering in this region.
Corresponding expression of the scattering efficiency has the
following form
\begin{equation}
 R\approx R^{(2)}(x,\varkappa ,q)=\frac{2^{4/3}
\pi }{\left|\beta \right|} {\rm Ai}\left[ \frac{\beta
^2-K^2}{2^{2/3}}  \right] \label{eq:R2}
\end{equation}
 Assuming the spectral density
$\left|\tilde n(\varkappa ,q,\Omega,x)\right|^2$  to vary with
$\varkappa$ slow enough and neglecting this variation when
performing the integration over $\varkappa $ which is correct for
distances from the cut-off $x_c-x>\ell_{cx}$ where $\ell_{cx}$ is
radial correlation length of the turbulence, we represent the
scattering efficiency in form~(\ref{eq:Sx}) and get the forward
scattering contribution
\begin{eqnarray}
\fl S_{FS}(x,q)=4\left[\rho
^4+\frac{c^2(\Lambda_0-\mathcal{R})^2}{\omega^2}\right]^{-1/2}
\frac{\left|f\left(-q/2\right)\right|^4}{k_x^2\left(x,-q/2\right)}
\left|\tilde
n\left(\frac{Lcq^2}{\omega}\frac{\Lambda_0-\mathcal{R}}{(\Lambda_0-\mathcal{R})^2+
\mathcal{P}^2},q,\Omega,x\right)\right|^2
\nonumber\\\lo{\times}\exp{\left\{-\frac{2(\rho
qL)^2}{(\Lambda_0-\mathcal{R})^2+ \mathcal{P}^2}
\left[\frac{ck_x\left(x,-q/2\right)}{\omega}\right]^2\right\}}
\label{eq:SFS:lin}
\end{eqnarray}
and backscattering one
\begin{equation}
\fl S_{BS}(x,q)= \left[\rho
^4+\frac{c^2(\Lambda_0-\mathcal{R})^2}{\omega^2}\right]^{-1/2}
\frac{\left|f\left(-q/2\right)\right|^4}{k_x^2\left(x,-q/2\right)}
\sum_{\pm }\left|\tilde n\left[\pm
2k_x\left(x,-q/2\right),q,\Omega,x\right]\right|^2
\label{eq:SBS:lin}
\end{equation}
This BS efficiency expression matches  when leaving  the cut-off
corresponding WKB formula~(\ref{eq:SBS}) obtained for the
arbitrary density profile whereas the FS
contribution~(\ref{eq:SFS:lin}) exceeds corresponding WKB
result~(\ref{eq:SFS}) by the factor of~$2$. Taking into account
the discussion of the previous section we can write approximate
formula describing the transition from~\eref{eq:SFS}
to~\eref{eq:SFS:lin} which happens when nonlinear correction to
the density profile decreases.
\begin{eqnarray}
\fl S_{FS}(x,q)\sim
4\frac{1+\gamma^2}{1+2\gamma^2}k^{-2}(x,\mathcal{K})\left|f\left(-\frac{q}{2}\right)\right|^4
\exp\left\{-\frac12\frac{\left[\rho
q\Lambda(x)\right]^2}{(\Lambda_0-\mathcal{R})^2+
\mathcal{P}^2}\right\}
\nonumber\\
 \lo{\times}\left[\rho
^4+\frac{c^2(\Lambda_0-\mathcal{R})^2}{\omega^2}\right]^{-1/2}\left|\tilde
n\left[\frac{q^2\Lambda(x)}{2k(x)}\frac{\Lambda_0-\mathcal{R}}{(\Lambda_0-\mathcal{R})^2+
\mathcal{P}^2},q,\Omega,x\right]\right|^2 \label{eq:SFS:approx}
\end{eqnarray}
where as above $\mathcal{P}=\omega\rho^2/c$ and
$\gamma=ck(x)/(\omega\rho)\sqrt{\alpha
c(\Lambda_0-\mathcal{R})/(12\omega)}$.

Expressions~(\ref{eq:SFS:lin}),~(\ref{eq:SBS:lin})
and~\eref{eq:SFS:approx} for the scattering efficiency describe
the transition from the ray tracing consideration to the case of
probing beam focusing to the cut-off both for the gaussian and
arbitrary antenna pattern. In the last case we determine
parameters $\rho$, $\mathcal{R}$ as:
\[
\rho^2=-\Re\frac{f''(\mathcal{K})}{f^2(\mathcal{K})},\qquad
\mathcal{R}=\Im\frac{\omega f''(\mathcal{K})}{cf^2(\mathcal{K})}
\]
In case when ray tracing approximation fails to hold, which can be
provided by large antenna beam width $\rho^2>c\Lambda_0/\omega$ or
by the focusing to the cut-off $\mathcal{R}\sim\Lambda_0$, the FS
contribution takes the following form
\begin{equation*}
\fl S_{FS}(x,q)\approx 4\left[\rho
k(x)\right]^{-2}\left|f\left(-\frac{q}{2}\right)\right|^4
\exp\left\{-\frac12 \left(\frac{cq}{\omega}\right)^2
\left[\frac{\Lambda(x)}{\rho}\right]^2\right\}\left|\tilde
n\left[0,q,\Omega,x\right]\right|^2
\end{equation*}

\Eref{eq:SBS} for BS contribution holds true in general situation
${\left(\mathcal{P}\lessgtr|\Lambda_0-\mathcal{R}|\right)}$ for
arbitrary antenna pattern providing the redefining of~$\rho$
and~$\mathcal{R}$ mentioned above is made.

The second important case to be considered is when the turbulence
is situated near the cut-off $K\lesssim 1$ where WKB scattering
efficiency has singularity. Main contribution to the scattering in
this location can be shown to be due to fluctuations with $|\beta
|<K$. It is given by
\begin{eqnarray*}
\fl R(x,\varkappa ,q)\sim 2\pi {\rm J_0}\left[ \frac12
\left(K^2-\beta ^2\right)\left|\beta -\theta \right| \right]
\rme^{-\epsilon\left(K^2-\beta ^2\right)/2 } -\pi\\
\lo{-}2\int_0^{\infty} \frac{\rmd\tau }{\tau }\,\sin \left[
\frac{\tau ^3}{6}+\frac{\tau }{2} \left(\beta ^2
-K^2\right)\right]
\end{eqnarray*}
This expression reveals a scattering efficiency maximum to be
situated near $K\approx 2.3$ which corresponds to $k(x)\sim
1.15/\ell$ or $x_c-x\sim 1.3 \ell$.
\begin{figure}
\begin{center}
\includegraphics[height=0.3\textheight]{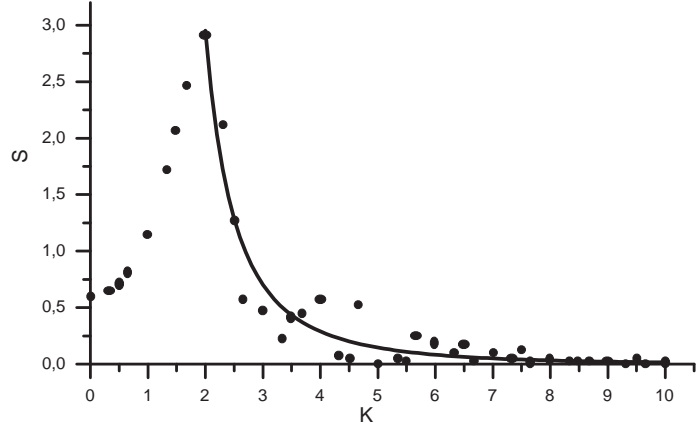}
\end{center}
\caption{\label{fig:Sx}Scattering efficiency calculated
numerically (\textbullet) and analytically (solid line) vs.
$K=2k_x(x,-q/2)\ell $.}
\end{figure}
A numerical calculation  of function $R(x,\varkappa ,q)$ with
following integration over $\varkappa $ confirms this result.
\Fref{fig:Sx} represents numerical results (black dots) in case of
short-scale fluctuations~$\ell_{cx}\ll\ell$. Here for the sake of
clarity the fluctuations spectral density is supposed to be
constant in~\eref{eq:Sx:COL} during the integration over
$\varkappa$.

 Additionally this computation
shows WKB formulae~(\ref{eq:SBS}),~(\ref{eq:SFS}) for scattering
efficiency to be valid up to the maximum providing $k_x$ in
denominators is replaced by
\begin{equation}
k_x\leftrightarrow k_x-0.5/\ell \label{eq:corr}
\end{equation}
(solid line in \fref{fig:Sx}). Oscillations near this solid line
correspond to the oscillating term omitted in~\eref{eq:R1} which
can be shown not to contribute essentially to the resulting
scattered signal magnitude due to averaging provided by slow
spatial variation of the fluctuation amplitude. Fast decay of $S$
for $k_x<0.5/\ell$ is caused by the probing wave field decrease in
the evanescent region.

In integration over $\varkappa$ in
expressions~(\ref{eq:Sx:osc}),~(\ref{eq:R1}),~(\ref{eq:R2}) we
neglected the influence of the fluctuation spectral density
$\left|\tilde n(\varkappa ,q,\Omega,x)\right|^2$. Now we consider
the case $\ell_{cx}\gg\ell$ when this approximation is not valid.
In this case of long-scale turbulence the backscattering
contribution is small in comparison with forward scattering one
and can be neglected
\begin{equation*}
 S(x)\approx\pi
\left(\frac{e^2}{m_e c^2 }\right)^2 \int_{-\infty}^{+\infty}
\frac{\rmd q}{2\pi }S_{FS}(x,q)
\end{equation*}
 According
to~(\ref{eq:Sx:COL}),~(\ref{eq:R1}) the scattering efficiency
takes the form
\begin{eqnarray}
\fl S_{FS}(x,q)\approx \frac{2c}{\omega\ell}
\left|f\left(-\frac{q}{2}\right)\right|^4 \rme^{-\epsilon K^2}
\int_{-\infty}^{+\infty}\rmd\beta\left|\tilde
n\left(\frac{\beta}{\ell} ,q,\Omega,x\right)\right|^2 {\rm
J_0}\left[ \frac12 K^2\left|\beta -\theta \right|
\right]\label{eq:Sx:long}
\end{eqnarray}
The characteristic scale of the spectral density variation with
$\beta=\varkappa\ell$ is $\ell/\ell_{cx}$. The corresponding scale
for the term ${\rm J_0}\left[1/2 K^2\left|\beta -\theta \right|
\right]$ is $\delta\beta\sim K^{-2}$. In evaluating last integral
in~\eref{eq:Sx:long} two cases can be distinguished. Far from the
cut-off at $x_c-x\gg\ell_{cx}$ the integral converges due to the
Bessel function at $\beta\sim\delta\beta\ll\ell/\ell_{cx}$. It
gives $k_x^{-2}$-behavior for the scattering efficiency which was
obtained above~(\ref{eq:SFS:lin}).

In the opposite case $x_c-x\lesssim\ell_{cx}$ the integral
in~\eref{eq:Sx:long} converges due to the turbulence spectrum at
$\beta\sim\ell/\ell_{cx}$ leading to the saturation of the
singularity $k_x^{-2}$. This saturation is described analytically
in the case of  gaussian fluctuation spectral density
\[
\left|\tilde n\left(\varkappa
,q,\Omega\right)\right|^2=2\sqrt{\pi}\ell_{cx}\left|\tilde
n\left(q,\Omega\right)\right|^2\rme^{-\ell_{cx}^2\varkappa^2}
\]
which allows the integral over $\beta$ in~\eref{eq:Sx:long} to be
calculated  exactly
\begin{eqnarray*}
\fl \int_{-\infty}^{+\infty}\rmd\beta\left|\tilde
n\left(\frac{\beta}{\ell} ,q,\Omega,x\right)\right|^2 {\rm
J_0}\left[ \frac12 K^2\left|\beta -\theta \right|
\right]\\\lo{\approx} 2\pi\ell\left|\tilde
n\left(q,\Omega,x\right)\right|^2\exp\left(-\frac{K^4\ell^2}{32\ell_{cx}^2}
\right){\rm I_0}\left(\frac{K^4\ell^2}{32\ell_{cx}^2}\right)
\end{eqnarray*}
where ${\rm I_0}$ is modified Bessel function.
\begin{figure}
\begin{center}
\includegraphics[height=0.3\textheight]{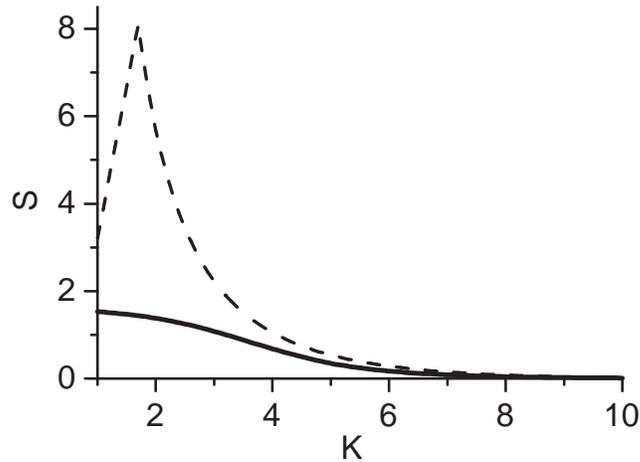}
\end{center}
\caption{\label{fig:satur}Approximation of scattering efficiency
for $\ell_{cx}\gg\ell$ (solid line $\ell_{cx}=2$~cm, $\ell
=0.5$~cm) and $\ell_{cx}\ll\ell$ (dashed line) vs.
$K=2k_x(x,-q/2)\ell $.}
\end{figure}

The difference in scattering efficiency behavior for small-scale
$(\ell_{cx}\ll\ell)$ and long-scale turbulence
$(\ell_{cx}\gg\ell)$ is illustrated by~\fref{fig:satur}. In our
case the transition to the plateau occurs in the point $K^2\sim
4\ell_{cx}/\ell$ which gives
\[
x_c-x\sim \ell_{cx}
\]

To summarize, in the case $\ell_{cx}\ll\ell$ in approaching the
cut-off the scattering efficiency grows, has maximum in the point
$x_c-x\sim 1.3 \ell$ and diminishes when $x_c-x<\ell$ due to
probing wave field decrease in the evanescent region. When
$\ell_{cx}\gg\ell$ the scattering efficiency has a plateau in the
region $x_c-x\lesssim \ell_{cx}$.
\section{Estimation of experiment locality}
\label{sec:Num} On performing the correction~(\ref{eq:corr}) the
WKB-formulae~(\ref{eq:SBS}),~(\ref{eq:SFS}) can be used up to the
cut-off vicinity. We consider simple model illustrating main
properties of the scattering efficiency obtained. Geometrical
parameters taken correspond to Tore Supra experiments~\cite{Zou},
where cut-off being situated in the antenna near-field zone
($\omega/c\sim 12.6\mbox{ cm}^{-1}$, $\rho\sim 14\mbox{ cm}$,
distance to the cut-off $L\sim 20\mbox{ cm}$).
\begin{figure}
\begin{center}
\includegraphics[width=\textwidth]{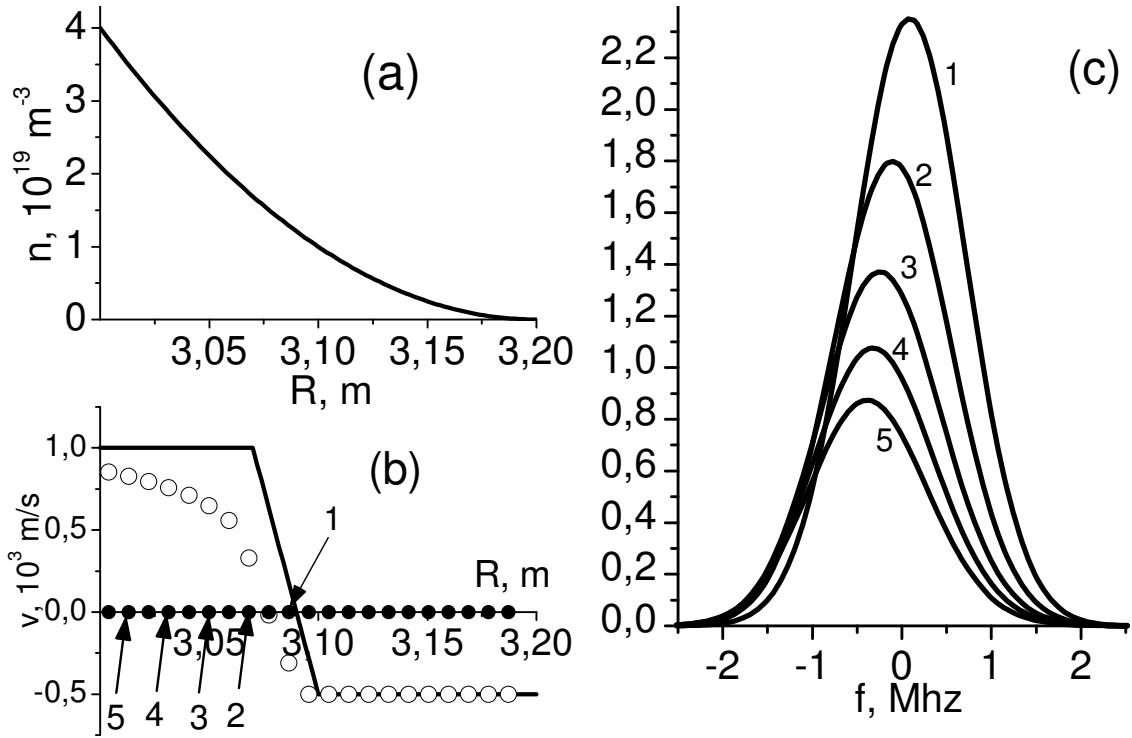}
\end{center}
\caption{\label{fig:Ex1} Signal spectrum evolution. (a)~Assumed
density profile. Horizontal axe is equivalent to  the major
radius. Probing is performed from right-hand side. (b)~Poloidal
velocity profile (solid line), cut-off positions~(\textbullet) and
velocity estimated using frequency spectrum shift~($\circ$).
(c)~Signal spectrum corresponding to the cut-off positions marked
on~(b).}
\end{figure}
Let us assume bent down plasma density profile~(\fref{fig:Ex1}(a))
similar to observed in Tore Supra~\cite{Clairet} and step-like
plasma poloidal velocity distribution~(\fref{fig:Ex1}(b)). For the
sake of simplicity we assume the turbulence level to be uniform
(which looks like real situation when relative turbulence
amplitude $\delta n/n$ increases in approaching plasma periphery)
and the wavenumber  spectra to be gaussian.

The probing is performed at different frequencies and therefore
with different cut-off positions. We assume probing antenna to
provide constant tilt angle $\vartheta=11.5^\circ$ and will
calculate corresponding probing wave poloidal wavenumber
$\mathcal{K}=\omega/c\sin\vartheta$ for each frequency.

We take into account scattered signal frequency shift due to the
Doppler effect in the final expression for the scattering
efficiency therefore assuming arising additional dependence on
radial coordinate $\left|\tilde
n\left[\dots,\dots,\Omega-qv(x)\right]\right|^2$ to be slow
enough. We perform the integration over $q$
in~\eref{eq:SBS:lin},~\eref{eq:SFS:approx} assuming the antenna
pattern~(\ref{eq:gauss}) to be wide enough to determine the
behavior of integrand and calculate spectral power
density~(\ref{eq:ps}) of the registered signal.  Spectra obtained
are represented in \fref{fig:Ex1}(c).

Turbulence correlation length was taken small $\ell _c\sim
0.1$~cm, Airy length being $\ell \sim 0.5$~cm. So the turbulence
spectrum did not improve enough the locality and one can see that
contribution of long area with $v=-0.5\cdot 10^5$~cm/s (see
\fref{fig:Ex1}(b)) is essential for spectra~$1$--$3$
(\fref{fig:Ex1}(c)). It can be seen that these spectra Doppler
shifts do not accord the velocity in corresponding cut-off
positions (see \fref{fig:Ex1}(d)). It is necessary to move
cut-off deep inside the plasma (spectra~$4$, $5$) to provide the
dominance of the region behind the velocity step point. Thus poor
locality of Doppler reflectometry associated with bend down
density profile in this situation can obscure the velocity
distribution.
\begin{figure}
\begin{center}
\includegraphics[width=\textwidth]{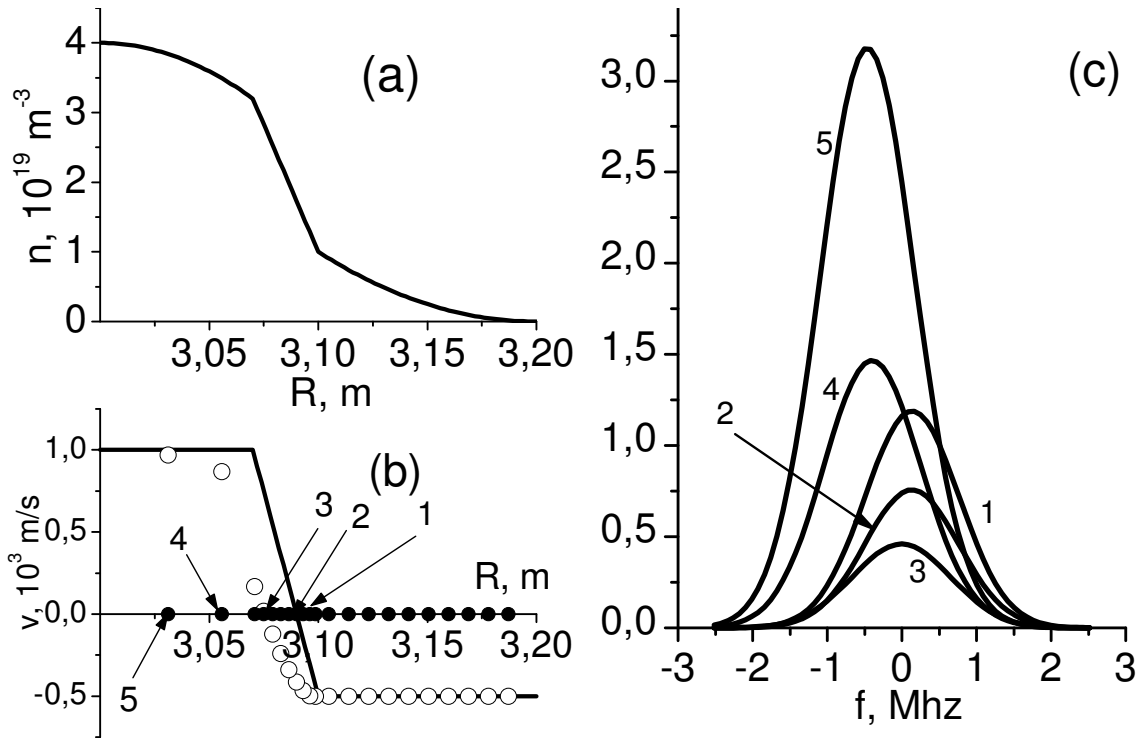}
\end{center}
\caption{\label{fig:Ex2} Signal spectrum evolution. (a)~Assumed
density profile. Horizontal axe is equivalent to  the major
radius. Probing is performed from right-hand side. (b)~Poloidal
velocity profile (solid line), cut-off positions~(\textbullet) and
velocity estimated using frequency spectrum shift~($\circ$).
(c)~Signal spectrum corresponding to the cut-off positions marked
on~(b).}
\end{figure}

Diagnostics localization dependence on plasma density profile is
also demonstrated by the next example simulating a transport
barrier. We consider plasma density profile~(\fref{fig:Ex2}(a))
bent down in plasma periphery and bent up in the core. The
poloidal velocity profile has high gradient in the ``barrier''
region~(\fref{fig:Ex2}(b)). One can see that up to cut-off
position~3 frequency spectrum shift (see spectra~$1$--$3$,
\fref{fig:Ex2}(c)) accords to negative poloidal rotation velocity
corresponding to the plasma periphery. But on crossing the twist
point by the cut-off position the signal grows and the frequency
shift changes to that corresponding to the local value of the
velocity in the cut-off region (cf. spectra~$4$, $5$,
\fref{fig:Ex2}(c) and \fref{fig:Ex2}(d)). This illustrates the
better locality of the method when used on bent up density
profile.

\begin{figure}
\begin{center}
\includegraphics[width=\textwidth]{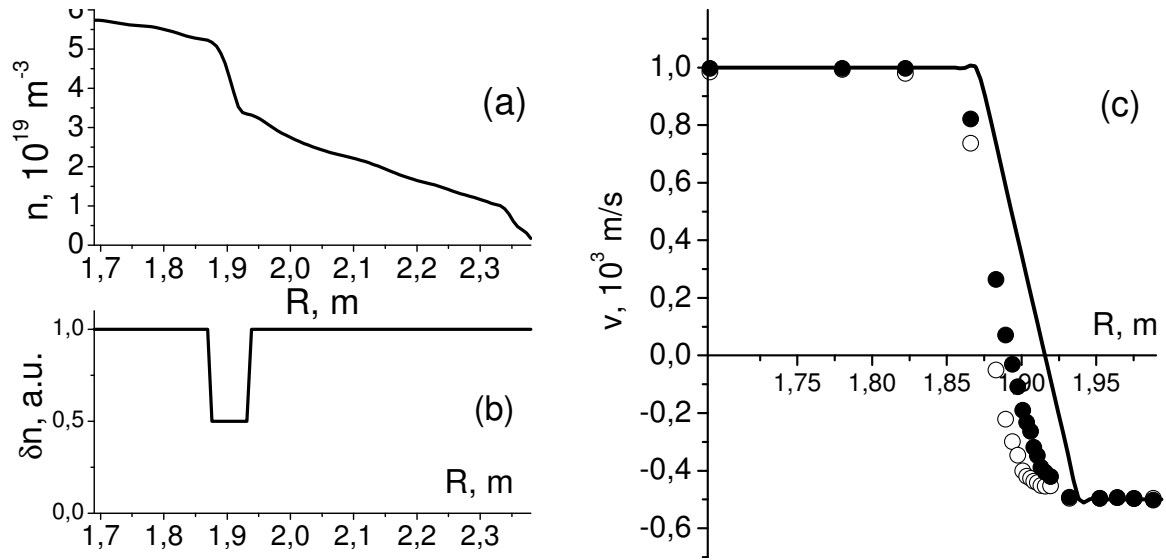}
\end{center}
\caption{\label{fig:Focus} Antenna focusing influence. (a)~DIII-D
density profile~\cite{Doyle}. (b)~Turbulence amplitude assumed.
(c)~Poloidal velocity profile (solid line), and velocity estimated
using Doppler reflectometry signal frequency spectrum shift:
\textbullet---using antenna focusing, $\circ$---without focusing.}
\end{figure}

To illustrate antenna focusing influence we consider the density
profile of \mbox{DIII-D} tokamak plasma with internal transport
barrier (\fref{fig:Focus}(a))~\cite{Doyle}. Here we take into
account the distance between antenna and the plasma, which was
assumed to be equal~$1$~m, and suppose that the probing is
performed with narrow antenna beam ($\rho\sim1$~cm) to provide
condition~(\ref{eq:condR}) to be satisfied. Besides that we take
into account the turbulence suppression in the barrier region
(see~\fref{fig:Focus}(b)).
Despite the fact that density profile in the barrier region is
favorable for the diagnostics, antenna focusing makes the spectrum
shift more adequate to the behavior of plasma velocity in the
cut-off.

\section{Discussion}
\label{sec:Disc}%
First of all we discuss the frameworks of approximations used. In
this paper we consider Doppler reflectometry in slab plasma
geometry. As it was mentioned in \sref{sec:Intro} this model is
reliable for large vertically elongated plasma. The effects of
cylindricalness become important when the probing beam width in
the cut-off vicinity is comparable with the cut-off surface
curvature radius which takes place in small toroidal devices or in
case of probing of plasma central regions. These cases of
essentially cylindrical plasma geometry were considered
numerically in~\cite{Lin00} and analytically (for specific plasma
density profiles) in~\cite{Bruskin02}. In these papers rather
obvious conclusion was obtained that plasma poloidal curvature
enhances the diagnostics sensitivity to the fluctuations with high
poloidal wavenumbers. Another cylindrical geometry effect was
investigated numerically in~\cite{Bulanin03} where strong
influence of the plasma cylindricalness on the diagnostics
locality was demonstrated.

Essential can be the plasma poloidal curvature influence on the
focused antenna beam. In the present paper it was neglected, which
is correct, when the focus radial shift due to refraction
associated with the plasma curvature is less than radial
fluctuation correlation length~$\ell_{cx}$. This criterion can be
represented in form
\[
2\Lambda_0\frac{x_c}{r_0}<\ell_{cx}
\]
Here $\Lambda_0$ is the ray trajectory length from the antenna to
cut-off, which was defined above, $x_c$ is the distance between
the antenna and the cut-off and $r_0$ is the plasma radius in
poloidal plane. Roughly this criterion can be approximated as
$x_c^2/r_0\lesssim\ell_{cx}$. When it fails to hold to provide the
focusing to the cut-off the probing wavefront curvature radius
should be calculated taking into account plasma poloidal
curvature. All the mentioned effects will be taken into account in
separate paper of the authors.

Additionally it should be noted that 2D plasma geometry effects
like cylindricalness are not a priory significant in reflectometry
(see e.g.~\cite{GusPopovEPS,Valeo02,Kramer}, where it was
demonstrated both analytically~\cite{GusPopovEPS} and
numerically~\cite{Valeo02,Kramer} that 2D theory predictions for
radial correlation reflectometry are similar to conclusions of
simple 1D model).

The second essential assumption made in the paper is associated
with linear character of the scattering signal formation.
According to~\cite{GusPopov} this approximation is correct when
following criterion is satisfied

\begin{equation} \frac{\delta
n^2}{n_c^2}\,\frac{\omega^2x_c\ell_{cx}}{c^2}\,\ln\frac{x_c}{\ell_{cx}}\ll
1 \label{eq:crit:lin}
\end{equation}
Here $\delta n$ is r.m.s amplitude of the turbulence and $n_c$
denotes the density in the cut-off. It can be seen
that~\eref{eq:crit:lin} can be violated in case of high
fluctuation amplitude or long trajectory length. In this case
probing wave multi-scattering should be taken into account. This
situation can be treated analytically by the procedure using
in~\cite{GusPopov} and it will be done in the paper by the authors
which is now under preparation for submission.

For diagnostics results interpretation it is important to
distinguish linear and nonlinear situation. To do that
experimentally one can use additional acquisition antenna (antenna
2 in \fref{fig:Scheme}) which receives the wave reflected from the
cut-off. If the specular component persists in the frequency
spectrum measured by this additional antenna the distortions of
the probing wave in propagation are weak and we deal with linear
situation of single-scattering. In the opposite case when the line
at probing frequency is lost in the broadened reflection spectrum
these distortions due to propagation in turbulent plasma lead to
extinction of the specular component, which indicates transition
to nonlinear small-angle multi-scattering regime. This way of
experimental confirmation seems to be reliable but needs
additional access to plasma. If it is impossible some information
can be obtained from the form of Doppler reflectometry spectrum
received  by antenna 1 in \fref{fig:Scheme}. If the frequency
spectrum width is consistent with estimation made based on the
antenna pattern width $(\delta\omega\sim2\sqrt2v/\rho)$ one can
conclude that the signal most likely resulted from
single-scattering.

The scattering efficiency introduced and analytically
obtained~(\ref{eq:SBS}),~(\ref{eq:SFS}) in the paper reveals main
similarities and differences of conventional reflectometry and
Doppler technique. In the both methods scattering signal is
proportional to reversed square of probing
wavenumber~\cite{GusPopovEPS}. That provides the diagnostics with
spatial localization, which can be rather poor in case of
unfavorable density profile.

The back-scattering signal component formation  is similar for
tilted probing, which is performed in Doppler technique and normal
probing which is specific for conventional method. In particular,
in both cases the probing wave focusing to the cut-off, according
to~(\ref{eq:SBS}), underlines the cut-off region contribution and
should improve the diagnostics localization. It is worth
mentioning that utilization of non-slab probing wave fronts have
been already discussed in~\cite{Hirsch01,Mazzu,Hirsch04}, however
in these papers the front curvature  was chosen close to the cut
off surface curvature in order to improve the Doppler
reflectometry wave number resolution~\cite{Hirsch01,Hirsch04}, or
to reduce the 2D interference effects in the signals reflected
from different parts of corrugated cut off surface~\cite{Mazzu}.
Finally  it should be noted that 2D-focusing onto the cut-off
surface, which can be easily realized in experiment,  can provide
even better localization than 1D-focusing  only possible within 2D
model considered in the present paper.

Essential are the peculiarities of FS efficiency~(\ref{eq:SFS}).
 Both for Doppler and conventional reflectometry the cut-off
contribution to the FS component is received by the most favorable
part of the antenna diagram $k_y=-q/2$. However the suppression of
the signal coming from the plasma volume in the case of
conventional reflectometry takes place only for fluctuations
satisfying condition $q\rho\gg1$, at which the cut off
contribution is suppressed as well. For longer poloidal scales
$q^{-1}\ge\rho$ the suppression is not efficient and additional
localization is not possible. On contrary for Doppler
reflectometry due to the tilted probing the FS component of the
signal, coming from the plasma volume is suppressed providing the
following condition is fulfilled
\[
\frac{\rho \mathcal{K}\Lambda_0}{\Lambda_0-\mathcal{R}}\gg1
\]
It can be easily achieved by large enough antenna tilt angle or
beam width, or by the focusing to the cut-off, thus making the FS
contribution  extremely localized to the cut-off.

Finally we discuss the wavenumber resolution of the Doppler
reflectometry. The scattering efficiency
obtained~(\ref{eq:SBS}),~(\ref{eq:SBS}) demonstrates that
diagnostics possess practically no radial wavenumber resolution
due to the fact that scattering signal is an integral over radial
wavenumbers and small $\varkappa$ are pronounced with weight
function $1/\varkappa$. Poloidal wavenumber resolution can be
easily estimated and is determined by antenna pattern width. For
BS contribution it can be represented as
\[
\delta q\sim\frac{\sqrt{2}}\rho
\]
FS poloidal wavenumber resolution can be $\sqrt2$ times worse.
\section{Conclusion}
\label{sec:Concl} In the present paper the Doppler reflectometry
spatial and wavenumber resolution is analyzed in the framework of
the linear Born approximation in slab plasma model. The results
obtained provide realistic description of Doppler reflectometry
experiment in large elongated plasma at low level of density
perturbation.

Explicit expression for the backscattering spectrum is obtained in
terms of wavenumber and frequency spectra of turbulence assumed to
be statistically inhomogeneous in radial direction. The treatment
is performed for arbitrary density profile and antenna pattern
taking into account diffraction effects. In agreement
with~\cite{Hirsch01} it is demonstrated that the signal consists
of contributions of back and forward scattering in radial
direction, which take place both before and after the reflection
of the probing wave in the turning point. Similar to the
traditional fluctuation reflectometry~\cite{GusPopovEPS} the
scattering efficiency for both back and forward scattering is
shown to be inverse proportional to the square of radial
wavenumber of the probing wave at the fluctuation location thus
making the spatial resolution of diagnostics sensitive to the
density profile. It is shown that additional localization is
provided in case of forward scattering in the radial direction by
the antenna diagram and in case of backscattering by the fact that
the turbulence spectrum is suppressed  at high radial wavenumbers.
The improvement of the diagnostics locality by probing beam
focusing onto the cut-off surface is proposed described as well.

It is demonstrated that analytical expressions obtained can be
easily used for fast interpretation of Doppler reflectometry data
in particular for estimation of this diagnostics locality. They
can as well serve for benchmarking and testing of full wave
numerical codes developed for interpretation of conventional
fluctuation reflectometry
data~\cite{Lin00,Bruskin02,Bulanin03,Valeo02,Kramer,GusLeclert},
should the authors consider the application of these codes to the
field of Doppler reflectometry.

The magnetic surfaces curvature, if important, can be accounted
for within the same theoretical approach applied to cylindrical
plasma geometry. The nonlinear effects in Doppler reflectometry
spectra formation becoming significant in large devices and at
high density perturbation level can be described within WKB
approximation in the way similar to one used in~\cite{GusPopov}.

\ack We would like to thank Dr.~V.V.~Bulanin (St.-Petersburg State
Polytechnical University) who drew our attention to the importance
of the probing wave front curvature for the diagnostic
performance.

This paper was subsidized by RFBR grants 02-02-17589, 04-02-16534,
 State support of leading scientific schools program
(project no.~2159.2003.2), INTAS grant 01-2056 and NWO-RFBR grant
047.009.009. One of the authors (A.V.S.) is grateful to the
Dynasty foundation for supporting his research.
\appendix
\section{}
\setcounter{section}{1}
\label{sec:App} Here we estimate additional
contribution to the forward scattering efficiency~(\ref{eq:SFS}),
which can be provided by singularities of the term $\left|\Delta
(\varkappa, q)\right|^2/\left|\varkappa \right|$
in~\eref{eq:Sx:osc}. This term possesses four branching points
which in ray tracing case
 are determined by equations
\[
\varkappa_{1-4}=\pm\frac{q^2}{\Lambda_0-\mathcal{R}}
\left[\frac{cL_*\Lambda_*}{\omega
k\left[x_*\left(\varkappa_{1-4}\right)\right]}\right]^{1/2}
\left(1\pm\frac{\rmi\omega\rho^2}{2c(\Lambda_0-\mathcal{R})}\right)
\]
where $L_*$,~$\Lambda_*$ are taken in the scattering point
$x_*\left(\varkappa_{1-4}\right)$ corresponding to
$\varkappa_{1-4}$.
 The branching point contribution is especially large in
the cut-off vicinity where $k^2(x)=\omega^2/c^2(x_c-x)/L_*$ and
\begin{equation*}
\Lambda(x)=\frac{2cL_*}{\omega}k(x)
\end{equation*}
 The real part
of right branching points $\varkappa_b\equiv\Re\varkappa_{1,4}$ in
this case coincides with the stationary phase point
$\varkappa^*\equiv\varkappa_{m,-m}^*$ which gives the main
contribution to the FS efficiency. The analysis in this case
becomes complicated and inaccurate in WKB approximation.  More
rigorous approach to this case will be developed below
in~\sref{sec:Airy} taking into account that close to the cut-off,
where density profile can be supposed linear, accurate solutions
of equation~(\ref{eq:W}) are available.

In general case these branching points are situated far from the
stationary point, so that their contribution to the
integral~(\ref{eq:Sx:osc}) is a quickly oscillating function of
$x$ and $q$ and therefore is negligible. This is easy to show
already in the case of profile slightly different from linear when
branching point $\varkappa_b$ is not so far from the stationary
one to allow us to decompose
\[
x_*(\varkappa_b)\approx a+x_*'(\varkappa_b-\varkappa^*)
\]
To estimate the contribution of branching point we perform the
integration over $\chi=\varkappa-\varkappa_b$ within vicinity of
the branching point taking into account that
\[
\frac{\left|\Delta (\varkappa, q)\right|^2}{\left|\varkappa
\right|}\approx\frac{\omega}{c(\Lambda_0-\mathcal{R})}\left[\chi^2+
\left(\frac{\omega\rho^2\varkappa_b}{2c(\Lambda_0-\mathcal{R})}\right)^2\right]^{-1/2}
\]
The phase in~\eref{eq:Sx:osc} takes the form
\[
\frac{\rmi[x-x_*(\varkappa,q)]^2}{2x_{*\varkappa}'}=-\frac{\rmi\omega
k^2(x)\ell_*^6}{c(\Lambda_0-\mathcal{R})}\left[1-
\left(\frac{\omega\Lambda_*}{2cL_*k(x)}\right)^2\right]q^2 -2\rmi
k^2\ell_*^2\chi
\]
This representation allows us to perform the integration over
$\chi$ and estimate the integral over $q$ in~\eref{eq:Sx:osc}. The
ratio between the contributions of the branching and stationary
points takes the form
\[
\frac{S_b(x)}{S_{FS}(x)}\sim\rho\sqrt\frac{\omega}{c(\Lambda_0-\mathcal{R})}
\left|1-\left(\frac{\omega\Lambda_*}{2cL_*k(x)}\right)^2\right|^{-1/2}
\]
The factor on right-hand side of this equation can be estimated as
\[
\left[1-\left(\frac{\omega\Lambda_*}{2cL_*k(x)}\right)^2\right]^{-1/2}
\sim \frac{\omega}{ck(x)}\sqrt\frac{12}{\alpha}
\]
where $\alpha=\left.\left|L_*^2/n_e(x_c)\cdot\rmd^2n_e(x)/\rmd
x^2\right|\right|_{x=x_c}$ characterizes the nonlinearity  of the
density profile. This leads to the
condition~(\ref{eq:crit:branch}), when contribution in question
can be neglected.
\end{document}